\documentclass[prb, showpacs, twocolumn]{revtex4}
\usepackage[english]{babel}
\usepackage[latin1]{inputenc}
\newcommand{\mrm}{\mathrm}
\newcommand{\mb}{\mbox}
\newcommand{\sss}{\scriptstyle}

\usepackage{graphicx}
\usepackage{dcolumn}
\usepackage{bm}

\begin{document}


\title{Clustering of vacancy defects in high-purity semi-insulating SiC}

\author{R. Aavikko} \email{Reino.Aavikko@hut.fi}
\author{K. Saarinen}\thanks{Deceased in December 2005}
\author{F. Tuomisto}
 \affiliation{Laboratory of Physics, Helsinki University of Technology,\\ 02015-HUT, Espoo, Finland}

\author{B. Magnusson}
 \altaffiliation[Also at ]{Norstel Ab, Sweden}
\author{N.T. Son}
\author{E. Janzén}
\affiliation{Department of Physics and Measurement Technology, \\Linköping University, SE-581 83, Linköping, Sweden}

\date{\today}

\begin{abstract}

Positron lifetime spectroscopy was used to study native vacancy defects in semi-insulating silicon carbide. The material is shown to contain (i) vacancy clusters consisting of 4--5 missing atoms and (ii) Si vacancy related negatively charged defects.  The total open volume bound to the clusters anticorrelates with the electrical resistivity both in as-grown and annealed material. Our results suggest that Si vacancy related complexes compensate electrically the as-grown material, but migrate to increase the size of the clusters during annealing, leading to loss of resistivity.

\end{abstract}


\maketitle

\section{Introduction}
\label{sec:introduction}

Silicon carbide (SiC) is a promising semiconductor material for  high-temperature, high-power, high-frequency and radiation  resistant applications. Semi-insulating SiC has shown great potential as a substrate material for SiC- and III-nitride microwave technology. SiC substrates can be made semi-insulating by introducing deep levels (either impurities or intrinsic defects) to the material. High-purity semi-insulating SiC can be grown by the High-Temperature Chemical Vapor Deposition (HTCVD) technique~\cite{AlexMRS2000}, in which pure source gases are used.

Native vacancies have been observed in the material in electron paramagnetic resonance experiments,\cite{AlexMRS2000,zvanut,konovalov,son_03,SonECSCRM2002} but their role in the electrical compensation is unclear. Positron lifetime spectroscopy is a method for studying vacancy defects in solid materials.\cite{posspect, hautojarvi_93} Positrons are repelled by positive ion cores and tend to get trapped at vacancies. Since the electron density in the vacancies is lower than in the bulk, positron lifetime increases and its value indicates the open volume of the defect. Using the relative intensities of different lifetime components, one can  determine the concentrations of different vacancy defects.

During the last few years SiC has been extensively studied using positron spectroscopy (see e.g. Refs.~\onlinecite{arpiainen_02,bauer-kugelmann_97,brauer_JPCM_98,brauer_PRB2512_96,brauer_PRB3084_96,britton_01,dannefaer_95,henry_03,kawasuso_96,kawasuso_98,ling_00,mokrushin_91,polity_99,puff_95,rempel_95,staab_MSF_01,uedono_97}).
These studies often involve irradiated materials. In this paper we use positron lifetime spectroscopy to study as-grown bulk HTCVD 4H-SiC samples, grown under different conditions. We observe clusters of 4--5 vacancies, which grow in size even up to 30 missing atoms after high-temperature annealing. We detect also smaller open volume defects, which we attribute to negative $V_\mrm{Si}$ related complexes. Comparison with the results of resistivity measurements shows that the presence of vacancy clusters anticorrelates with the semi-insulating properties of the material. We suggest that the clusters act as neutral agglomeration centers for the negative vacancy defects causing the high resistivity.

The paper is organized as follows. The details of the positron experiments and theoretical calculations are given in Sec.~\ref{sec:method}. The positron results in the as-grown and high-temperature annealed SiC samples are given and interpreted in Sec.~\ref{sec:results}. In Sec.~\ref{sec:discussion} we present the identification of the vacancy defects in the light of theoretical calculations and earlier investigations, and discuss the vacancy concentrations and the effect of vacancy defects on the electrical compensation. Finally Sec.~\ref{sec:conclusions} concludes the paper.

\section{Method} \label{sec:method}

\subsection{Experimental details}

We measured over 20 samples grown by the HTCVD method above 1900 $^\circ$C in either hydrocarbon rich (type A) or poor (type B) conditions (examples shown in table~\ref{tab:samples}). The studied samples are undoped and their impurity levels are $< 10^{16}~\mrm{cm}^{-3}$ as measured by secondary ion mass spectrometry. Generally all as-grown samples are insulating but sample A1 shows weak n-type conductivity. Post-growth annealings were performed at \mb{1600 $^\circ$C} in H$_\mrm{2}$ ambient in a CVD reactor. After annealing the resistivity of A-type samples decreased and current-voltage measurements indicated that the samples became more n-type. The type B remain semi-insulating also after annealing.

\begin{table}[t]
\caption{Summary of samples studied in this paper. 
Annealings have been performed at 1600 $^\circ$C in H$_\mrm{2}$ ambient. Type A samples are grown in hydrocarbon rich and type B samples in hydrocarbon poor environment. Resistivities are presented at \mb{300 K}.}
\begin{tabular}{llll@{~~~}l@{~~~}ll} 
\hline\hline
Sample &  C$_2$H$_6$& Ann. & Resistivity & \\
&&&$[\Omega~\mathrm{cm}]$ & \\
\hline
A1 & rich &0 h   & n-type              \\
A1 & rich &1 h   & n-type               \\
A2 & rich &0 h   & $2\times$$10^{9}$\\
A2 & rich &1 h   & $1\times$$10^8$ \\
A2 & rich &2 h   & $4\times$$10^4$(n) \\
A2 & rich &3 h   & $5\times$$10^2$(n) \\
B & poor  &0 h   & $>10^{10}$           \\
B & poor  &1 h   & $>10^{10}$ \\
\hline\hline
\end{tabular}
\label{tab:samples}
\end{table}

\begin{table*}[tb]
\caption{Samples, their annealing times (at 1600 $^\circ$C in H$_\mrm{2}$ ambient), resistivities (at 300 K), positron lifetime values (measured at 300 K), positron trapping rates, vacancy concentrations and cluster sizes. Type A samples are grown in hydrocarbon rich and type B samples in hydrocarbon poor environment. 
The positron results include the measured average positron lifetime $\tau_\mrm{ave}$, the fitted positron lifetime components $\tau_{\{1,2\}}$ and the intensity of the longer lifetime component $I_2$. The positron trapping rates to monovacancies $\kappa_1$ and to vacancy clusters $\kappa_2$, vacancy defect concentrations for $V_\mrm{Si}$-related defects $[V_\mrm{Si}]$, vacancy clusters $[V_\mrm{N}]$ and cluster sizes $N$ have been determined by the positron annihilation measurements presented. Note that the change of Fermi-Level in type A2 sample evidently changes the charge state (and thus the positron trapping coefficient) of the defects (calculated concentrations \emph{highlighted}). } \label{tab:data} \vspace{1mm}
\begin{tabular}{llccccccccc@{~~~}c@{~~~}ccccccc} 
\hline\hline
Sample,& Resistivity & $\tau_\mrm{ave}$ & $\tau_1$ & $\tau_2$ & $I_2$ &  $\kappa_1$ & $\kappa_2$ & \hspace{2mm} $[V_\mrm{Si}]$ & $[V_\mrm{N}]$ & \hspace{1mm} Cl. Size $N$ \vspace{0.5mm}\\
annealing \hspace{1mm}&$[\Omega~\mathrm{cm}]$&[ps] &[ps] &[ps] & \%  & \multicolumn{2}{c}{$[10^{9}~\mrm{s}^{-1}]$} & \multicolumn{2}{c}{  \hspace{1ex} $[10^{15}~\mrm{cm}^{-3}]$} & [at.] \vspace{0.5mm}\\
time&& \hspace{1mm} $\pm 0.5$ \hspace{1mm} & \hspace{1mm} $\pm 2$ \hspace{1mm} & \hspace{1mm} $\pm 10$ \hspace{1mm} & \hspace{1mm} $\pm 2$ \hspace{2mm} & \multicolumn{2}{c}{$\pm 20$ \%} & \multicolumn{2}{c}{$\pm 20$ \%} & $\pm 20$ \% \\
\hline
A1 & n-type           &  
191 & 168 & 284 & 20& 
 \hspace{1mm} 7.5 \hspace{1mm} & \hspace{1mm} 2.6 \hspace{1mm} & 
\hspace{2mm} 120   & 50 & 5    \\ 
A1, 1h & n-type               &
208 & 138 & 350 & 33 &
2.1 & 2.9 & 
\hspace{2mm} 30 & 28  & 10 \\
A2 & $2\times$$10^{9}$    & 
163 & 152 & 283 & 8 &
1.3 & 0.4 &  
\hspace{2mm}\emph{61}   & \emph{8}  & 5    \\
A2, 1h & $1\times$$10^8$      & 
183 & 146 & 406 & 14 &
1.4 & 0.9 &  
\hspace{2mm}\emph{66}   & \emph{6}  & 16   \\
A2, 2h & $4\times$$10^4$ (n)   & 
206 & 144 & 440 & 21 &
2.0 & 1.7 &   
\hspace{2mm}\emph{33}   & \emph{6}  & 27   \\
A2, 3h & $5\times$$10^2$ (n)   & 
216 & 144 & 451 & 24 &
2.4 & 2.1 &   
\hspace{2mm}\emph{39}   & \emph{6}  & 32   \\
B & $>10^{10}$           & 
164 & 150 & 261 & 13 &
1.3 & 0.6 &   
\hspace{2mm}61  & 15 & 4  \\
B, 1h & $>10^{10}$           & 
179 & 146 & 342 & 17 &
1.4 & 1.0 &   
\hspace{2mm}69   & 11  & 9  \\
\hline\hline
\end{tabular}
\label{tab:samples_results}
\end{table*}

The samples were measured in darkness between \mb{20--540 K} with analog and digital~\cite{nissila_03} positron lifetime spectrometers. The time resolutions of the spectrometers were 240 ps and 215 ps (FWHM), respectively. The digital spectrometer was used for room temperature measurements. 
The detectors were equipped with plastic scintillators and were positioned in face-to-face geometry in both cases. We used $10~\mu$Ci $^{22}$Na positron sources deposited on folded $1.5~\mu$m Al foil. Typically $1-2\times 10^6$ counts of annihilation events were collected to each lifetime spectrum. The lifetime spectrum was analyzed as the sum of exponential decay components  $\sum_i I_i \exp(-t/\tau_i)$, where the $\tau_i$ are the individual lifetime components and $I_i$ their intensities. The increase of the average positron lifetime $\tau_\mrm{ave} = \sum{\tau_i I_i}$ above the bulk lattice lifetime $\tau_\mrm{b}$ shows that vacancy defects are present in the material. The average lifetime $\tau_\mrm{ave}$ is insensitive to the decomposition procedure, and even as small a change as 1 ps can be reliably measured.

\begin{figure}[tb]
\centering{\includegraphics[angle=0,width=\linewidth]{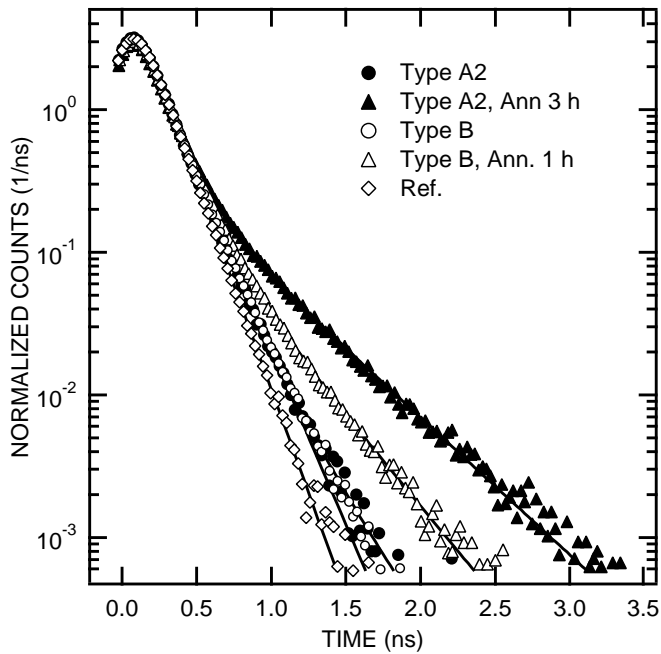}}
\caption{Positron lifetime spectra at 300 K. The solid lines are fits of sums of exponential decay components.}
\label{fig:LTSpectra}
\end{figure}

If the samples contain only one vacancy type positron traps with positron lifetime $\tau_D$, the positrons have two different states from which to annihilate, \emph{i.e.} bulk and defect. The longer experimental lifetime component will then be equal to that of the positron lifetime in the defect, \emph{i.e.} $\tau_2^\mrm{exp}=\tau_D$. Because of the positrons trapping away from the bulk at rate $\kappa$, the shorter experimental lifetime component becomes $\tau_1^\mrm{exp}=(\tau_\mrm{b}^{-1}+\kappa)^{-1}$. This effect can be used to test whether the one-trap model is sufficient to explain the measured data. For more details on the analysis, see e.g. refs.~\onlinecite{posspect, hautojarvi_93}.

\subsection{Theoretical calculations}
\label{sec:theorCalc}

We calculated  theoretically the positron lifetimes in vacancy clusters for the 4H polytype of SiC. For the positron states we used the conventional scheme with the local density approximation (LDA) for electron-positron correlation effects and the atomic superposition method in the numerical calculations.\cite{puska_83, puska_94} The positron annihilation rate $\lambda$ is

\begin{equation}
  \lambda=\tau^{-1} = \pi r_0^2 c \int d\mathbf{r} \left|\psi_+(\mathbf{r})\right|^2 n_{-}(\mathbf{r}) \gamma[n_-(\mathbf{r})],
\end{equation}

\noindent where $n_-$ is the electron density, $\psi_+$ the positron wave function, $r_0$ the classical electron radius, $c$ speed of light, and $\gamma$ the enhancement factor. We used a modified Boronski-Nieminen enhancement factor,\cite{boronski_86,puska_91} which takes into account lack of complete positron screening in semiconductors. The factor takes form

\begin{eqnarray}\label{eq:enhance}
  \gamma[n_-(\mathbf{r})] &=& 1+1.23 r_s+0.8295r_s^{3/2}-1.26r_s^2 \nonumber \\
&&+ 0.3286 r_s^{5/2}  + (1-1/\epsilon_\infty)r_s^3/6,
\end{eqnarray}

\noindent where $r_s$ is calculated from the electron density as $r_s=\sqrt[3]{3/(4\pi n_{-})}$. For the high-frequency dielectric constant in 4H polytype we use value $\epsilon_\infty=6.78$.~\cite{shaffer_71,mutschke_99}

We estimate the sizes of the observed vacancy clusters using these theoretical calculations. The calculated clusters are formed by removing Si-C pairs from the perfect lattice up to the size of 84 atoms. The atoms are removed according to their distance to the ''origin'' of the cluster (chosen to be the middle point between a Si-C bond, cubic position). The positron state is then solved in $480-2N$ atom supercells, where $N$ is the number of Si-C pairs removed. 

Because the atoms are simply removed from their ideal positions, the calculated clusters are non-relaxed. In general, the relaxation affects the open volume---and thus the positron lifetime of the clusters. Additionally, the presence of a trapped positron may alter the the configuration of the surrounding atoms. According to calculations, lattice relaxations have been found to affect the positron lifetimes in case of $V_\mrm{Si}$ in SiC by approximately 12\ldots15 ps and in $V_2$ by 4\ldots7 ps.~\cite{staab_MSF_01} Furthermore, the relative effect of relaxations in clusters have been found to diminish to insignificant level in clusters bigger than $V_4$ in silicon~\cite{staab_99} (note that the open volume of $V_4$ in Si is larger than that of $V_4$ in SiC).

\begin{figure}[tb]
\centering{\includegraphics[angle=0,width=\linewidth]{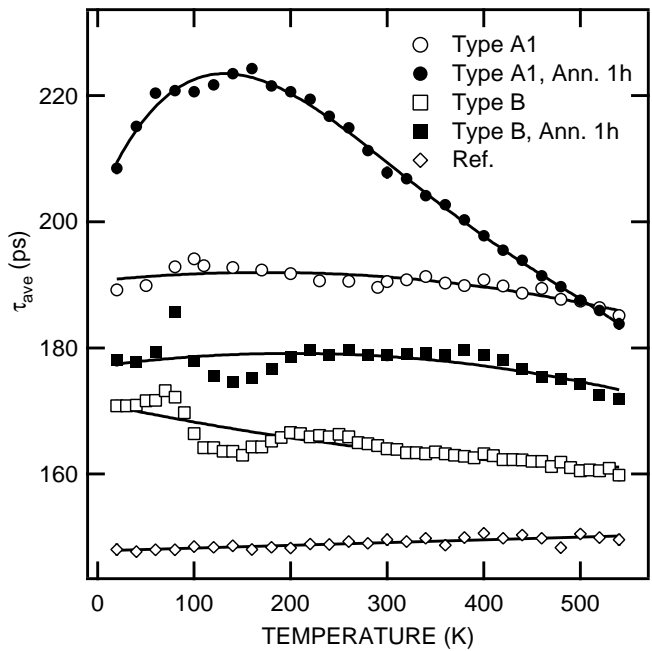}}
\caption{Average positron lifetime as a function of measurement temperature. The solid curves are used for calculation of $\tau_\mrm{1}^\mrm{\sss TEST}$ (see eq. \ref{eq:tautest}).}
\label{fig:LT_TEMP}
\end{figure}

\begin{figure}[tb]
\centering{\includegraphics[angle=0,width=\linewidth]{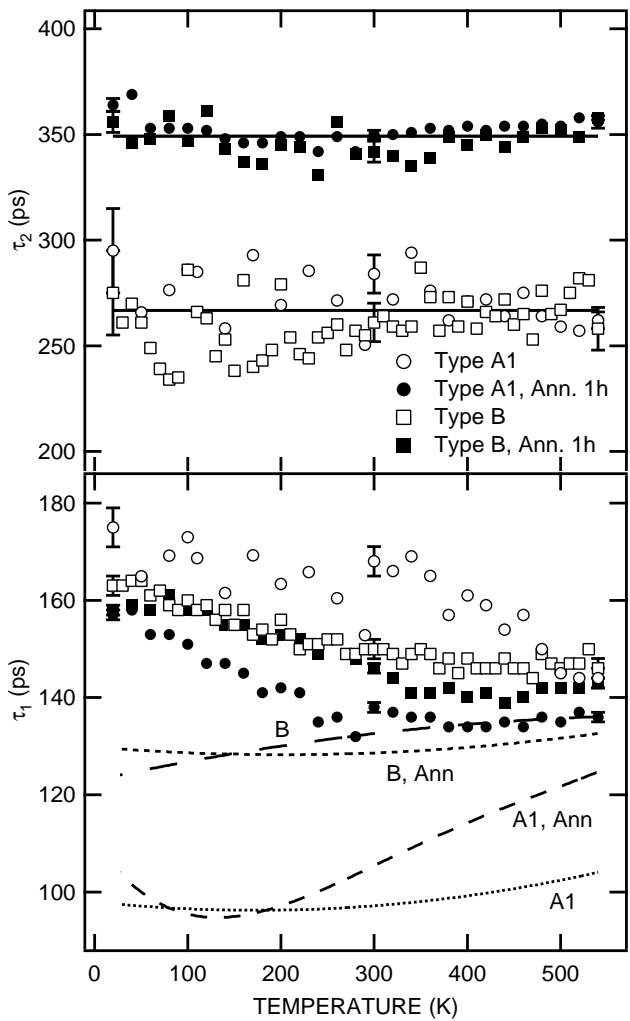}}
\caption{Positron lifetime components vs. measurement temperature. In the lower 
panel the lines present the parameter $\tau_\mrm{1}^\mrm{\sss TEST}$ (Eq.~\ref{eq:tautest}), which gives information on the number of different vacancy defect species in the samples.}
\label{fig:LT_comp}
\end{figure}

\section{Positron lifetime results}
\label{sec:results}

Positron lifetime spectra are shown in Fig~\ref{fig:LTSpectra}. The reference sample, p-type bulk SiC, shows only a single lifetime of 150 ps, which we attribute for the positron lifetime $\tau_\mrm{b}$ in the SiC lattice. All HTCVD samples have at least two lifetime components.

The lifetime spectra were decomposed into two components. Table~\ref{tab:data} presents the average positron lifetime at 300 K and the two separated components. The intensity of the longer component is also shown. The longer lifetime is 260-290 ps in the as-grown state and increases up to 450 ps after annealing. These lifetime values are typically associated to vacancy cluster consisting of more than two missing atoms.

The positron lifetime measurements as a function of temperature are shown in Fig.~\ref{fig:LT_TEMP}. We see that the average lifetime above 200 K is constant or decreases with temperature. This suggests  that negative vacancies are present in the samples, since positron trapping to negative vacancies decreases with temperature, whereas trapping to neutral vacancies is temperature independent.~\cite{posspect, hautojarvi_93} Below 200 K, especially well seen in the sample A1 Ann, the average lifetime starts to decrease, suggesting the presence of negative ions (residual impurities or intrinsic defects), which act as shallow traps for positrons. The samples B and B Ann have oscillations in $\tau_\mrm{ave}$ around 100 K, which may reflect some charge transitions in the shallow traps. However, the concentrations of the shallow traps are low compared to those of the vacancy defects, as the decrease of the average lifetime is modest compared to the difference between the average and bulk lifetimes.

Fig.~\ref{fig:LT_comp} shows the temperature dependence of the positron lifetime components. The positron lifetime at vacancy clusters $\tau_2$ is approximately constant (350 ps  or 270 ps) indicating positrons annihilating in a well-defined defect state in  in each sample. On the other hand, the shorter lifetime $\tau_1$ has a clear tendency to decrease from 160-175 ps at 20 K to 140-150 ps at 500 K. If only the vacancy clusters corresponding to $\tau_2$ were present in the samples, the lifetime $\tau_1$ in the lattice would be related to $\tau_\mrm{ave}$, $\tau_\mrm{b}$ and $\tau_2$ as

\begin{equation}\label{eq:tautest}
\tau_\mrm{1}=\tau_\mrm{1}^\mrm{\sss TEST} 
=\tau_\mrm{b} \left( \frac{\tau_2-\tau_\mrm{ave}}{\tau_2-\tau_\mrm{b}}\right) .
\end{equation}

The test lifetime $\tau^\mrm{TEST}$ calculated from the experimental values of $\tau_\mrm{ave}, \tau_\mrm{b}$ and $\tau_2$ varies between 95-137 ps and thus it is well below the experimental $\tau_1$ in all samples at any temperature. This means that the vacancy clusters corresponding to $\tau_2$ are not the only defects. There exist also other smaller vacancy defects, which create a lifetime component mixed into the experimental $\tau_1$ lifetime. The smaller defects are especially prominent at low temperatures and their concentration in the sample A1 grown under the hydrocarbon rich condition is high. The positron lifetime at the smaller vacancy defects is estimated to be above 170 ps  from the low-temperature part of the Fig.~\ref{fig:LT_comp}.
On the other hand, the $\tau_\mrm{ave}$ vs. T data in Fig.~\ref{fig:LT_TEMP} indicates that the defect-specific lifetime is below 220 ps.

\section{Discussion} \label{sec:discussion}

\subsection{Identification of the vacancy defects}

\subsubsection{Vacancy clusters}

The measured lifetime $\tau_\mrm{2}$ is longer than that determined earlier (typical values reported in literature shown in parenthesis, see more discussion about lifetime values at the end of this section) for the carbon vacancy $V_\mrm{C}$ ($\le$160 ps), 
the silicon vacancy $V_\mrm{Si}$ (180-210 ps), or divacancy (< 250 ps) in SiC.
The electron density in the defect is thus lower, indicating that the observed defects are vacancy clusters.

\begin{figure}[tb]
\centering{\includegraphics[angle=0,width=\linewidth]{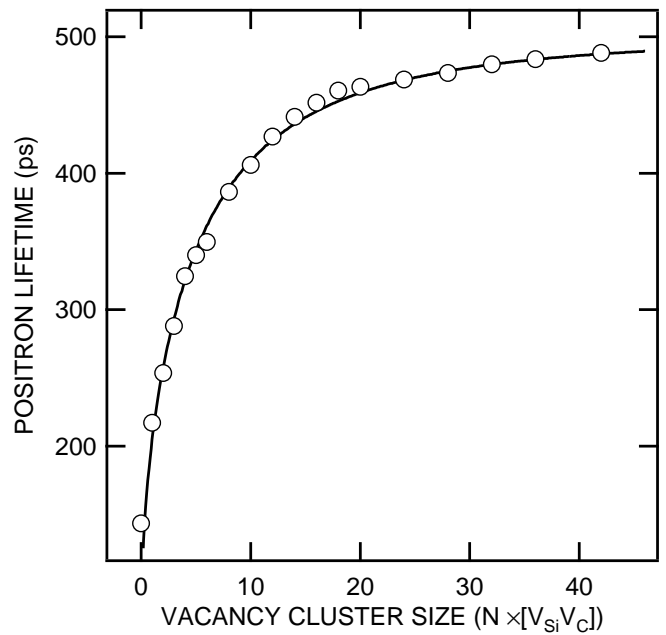}}
\caption{Calculated positron lifetimes in vacancy clusters in 4H SiC. The solid line is used for determining the cluster sizes from the measured $\tau_2$. It is worth noticing that the sensitivity of the positron lifetime on the vacancy cluster size is significantly reduced when the cluster size exceeds 10 missing Si--C pairs.}
\label{fig:clusterCalc}
\end{figure}

The calculated positron lifetimes in vacancy clusters are presented in figure~\ref{fig:clusterCalc}. The results obtained are similar to those reported earlier for 3C and 6H SiC.\cite{brauer_JPCM_98} Our calculations give for the positron lifetime in bulk SiC a value of $\tau_b = 143.7$~ps, in agreement with the measured value $\tau_b = 149.6$~ps. 
When using the theoretical calculations to determine the measured cluster sizes, we compare the differences between the bulk and defect lifetimes.

According to the theoretical calculations, a measured positron lifetime of 260 ps observed in as-grown samples corresponds to a vacancy cluster $V_\mrm{4}$ (2 Si--C molecules removed).  The lifetime of 350 ps, detected after annealing, is expected for an open volume of a cluster $V_\mrm{10}$. The estimated cluster sizes in different samples are presented in Table~\ref{tab:data}. We observe that the cluster sizes increase in the annealings from the size of roughly 5 missing Si--C pairs in the as-grown material up to 30 missing Si--C pairs in the sample annealed for 3 hours. These approximate values represent the averages of the open volume distribution of vacancy clusters. This may mean various different cluster sizes or possibly a "magic" cluster size, in which the number of dangling bonds are minimized, as shown in e.g. Si and GaAs.\cite{chadi_88,staab_99} Clustering of vacancies due to annealing has been previously reported in n-type 6H SiC after neutron- and ion irradiation.~\cite{mokrushin_91,anwand_MSF_01,anwand_ASS_02}

\subsubsection{Si vacancy related defects}
\label{sec:siVac}

The estimated lifetime of the positrons trapped at the smaller observed vacancy defects is $\tau_V = 195\pm 25$~ps range, \emph{i.e.} $\tau_V - \tau_b = 20 \ldots 70$~ps. In order to identify this defect, we need to consider the different lifetime values that have been associated with different aspects of SiC. Reported positron lifetimes (both experimental and theoretical) for bulk material range between 134 and 150 ps. The lifetime values associated with different vacancy defects vary clearly more.

For 4H polytype, bulk lifetime values including 141 ps~\cite{dannefaer_MSF_05}, 145 ps~\cite{bauer-kugelmann_97} 150 ps \cite{puff_95}, and for 3C-SiC 140 ps have been reported.~\cite{kawasuso_98} The most common polytype encountered in recent positron studies for SiC is 6H, for which several values for bulk lifetime have been proposed, in the range $136 \ldots 150$~ps. \cite{kawasuso_96,polity_99,dannefaer_95,rempel_95,henry_03,arpiainen_02,ling_00,puff_95} The theoretical calculations give lifetimes of 134 ps for 4H-SiC~\cite{staab_MSF_01}, 141 ps for 6H-SiC and 138 for 3C-SiC.~\cite{brauer_PRB2512_96} It should be noted that different calculation schemes (especially different enhancement factors) cause differences in the resulting absolute lifetimes, which becomes evident when comparing results from different studies.~\cite{staab_MSF_01} Somewhat similar situation applies to experimental values (e.g. due to differences in measurement geometry, energy windows or source corrections). Thus, in comparisons between different lifetime values obtained from different studies, we compare preferably the differences between determined lifetime values in defects and bulk ($\tau_V - \tau_b$), rather than focus solely on the absolute values of the lifetimes.

The studies of positrons trapping in vacancy-type defects are typically performed by making use of irradiation, and the identification of the defects is often based on comparing the measured and the theoretical positron lifetime values. The reported values for positron lifetimes in $V_\mrm{Si}$ based on irradiation experiments are typically around $\tau_V = 200$~ps, in the range $\tau_V - \tau_b = 14 \ldots 116$~ps.~\cite{rempel_95,kawasuso_98,dannefaer_95,henry_03,arpiainen_02,dannefaer_DRM_04} Theoretical calculations reported in the literature predict $\tau_V - \tau_b = 47 \ldots 64$~ps in 4H and 6H SiC \cite{brauer_PRB2512_96,staab_MSF_01} for $V_\mrm{Si}$. For $V_\mrm{C}$ values of $\tau_V - \tau_b =$ 8 and 16 ps have been reported~\cite{dannefaer_95,dannefaer_DRM_04} and theory gives $\tau_V - \tau_b = 6 \ldots 12$~ps.~\cite{brauer_PRB2512_96,staab_MSF_01}

In addition to the previous values, lifetime differences in the range $\tau_V - \tau_b = 65 \ldots 94$~ps have been reported in irradiated samples.~\cite{polity_99,rempel_95,uedono_97,brauer_PRB3084_96} The experimental values reported\cite{henry_03,arpiainen_02} $\tau_V - \tau_b = 80 \ldots 83$~ps for the divacancy ($V_\mrm{Si}V_\mrm{C}$) are in good agreement with the reported theoretical predictions $\tau_V - \tau_b = 73 \ldots 75$~ps for 4H~\cite{staab_MSF_01} and 6H~\cite{brauer_PRB2512_96} SiC. 

Vacancies are often detected also in as-grown materials. In many cases the lifetime values are somewhat longer than the values found after irradiation. Lifetime values mainly in the range $250 \ldots 350$~ps ($\tau_V - \tau_b= 100 \ldots 200$~ps) have been observed in as-grown SiC samples.~\cite{puff_95,dannefaer_95,kawasuso_96, britton_01} Also smaller values of $\tau_V - \tau_b= 50\ldots 75$~ps have been reported \cite{ling_00} and attributed to $V_\mrm{Si}$ and $V_\mrm{Si}V_\mrm{C}$. According both to our and earlier reported \cite{brauer_JPCM_98} theoretical calculations, a lifetime of 250~ps corresponds to open volume of at least similar size with that of $V_4$, \emph{i.e.} $(V_\mrm{Si}V_\mrm{C})_2$.

In the light of the lifetime values discussed previously, the possible candidates for the smaller vacancy defects observed here are thus monovacancies or monovacancy related complexes in the Si and C sublattices and small clusters with at most 2--3 missing atoms. Several arguments point in the direction that the observed defects are related to $V_\mrm{Si}$ rather than $V_\mrm{C}$. The main point is that, as shown above, the lifetime of the smaller vacancy defect is $\ge$170 ps, \emph{i.e.} above all presented estimates for the carbon vacancy $V_\mrm{C}$. In addition, the C vacancy in semi-insulating SiC could be positively charged according to theoretical calculations~\cite{torpo_01} and EPR experiments~\cite{SonECSCRM2002} and thus repulsive to positrons. Hence we attribute the defect responsible for the increase of $\tau_\mrm{1}$ at low temperature to the Si vacancy or a complex involving $V_\mrm{Si}$.  This conclusion is in agreement with the EPR and infrared absorption  results,\cite{SonECSCRM2002, AlexMRS2000, son_03} which suggest that samples similar to A1 contain $V_\mrm{Si}$, whereas samples of type B have positive C vacancies, and also $V_\mrm{Si}$ but less than in a sample like A1.  Absorption measurements~\cite{AlexMRS2000} further show that the concentration of $V_\mrm{Si}$-related center $T_\mrm{V2a}$~\cite{son_03} decreases in annealing of A-type samples, which correlates with the behavior of $\tau_\mrm{1}$ in Fig.~\ref{fig:LT_comp}. However, based on our positron measurements only, we cannot rule out the possibility of the defect being \emph{e.g.} the divacancy $V_\mrm{Si}V_\mrm{C}$.

It is interesting that the Si vacancy related defects are prominent at low temperatures, whereas the larger vacancy clusters dominate the positron lifetime spectrum at high measurement temperatures. Positron trapping at neutral defects is independent of temperature, whereas the negative defects become stronger positron traps at lower temperatures.~\cite{posspect, hautojarvi_93}  Hence, the temperature dependence of the lifetime components suggests that the observed Si vacancy related defects act as acceptors in SI HTCVD SiC, but the vacancy clusters are electrically neutral. In the annealed A1 type sample, however, the average positron lifetime decreases with the increasing temperature. This means evidently that a part of the clusters in the $n$-type material are negatively charged.

\subsection{Vacancy defect concentrations}

The vacancy defect concentrations can be estimated from the positron trapping model using the decomposed lifetimes and intensities.  In the calculation we assume that both vacancy clusters (CL) and vacancies in Si-sublattice (V) are present and the positron lifetimes at the vacancies in the Si-sublattice and in the bulk SiC are intermixed to the component $\tau_\mrm{1}$. The positron trapping  rates $\kappa$ are~\cite{posspect, hautojarvi_93}

\begin{eqnarray}
\kappa_\mrm{V} &=& \frac{\tau_\mrm{1}(\tau^{-1}_\mrm{b}-I_\mrm{2}\tau^{-1}_\mrm{2})-I_\mrm{1}}{(\tau_\mrm{V}-\tau_\mrm{1})},\label{eq:kappa1}\\
\kappa_\mrm{CL} &=& \frac{I_\mrm{2}}{I_\mrm{1}}(\tau^{-1}_\mrm{b} - \tau^{-1}_\mrm{CL} + \kappa_\mrm{V}).\label{eq:kappa2}
\end{eqnarray}

\noindent The positron lifetime for the vacancy cluster is taken directly from the decomposition as $\tau_\mrm{CL}=\tau_\mrm{2}$. For the positron lifetime at the Si vacancy related defects we use 
$\tau_\mrm{V}=195~\mathrm{ps}$, which is in the middle of our range determined for the lifetime of the smaller defect. Values close to this are also often attributed for the positron lifetime in $V_\mrm{Si}$ (see section~\ref{sec:siVac}).
The defect concentrations can be obtained from the trapping rates $\kappa$ as $c = \kappa~N_\mrm{at}/\mu$. Here $N_\mrm{at}=9.64\times10^{22}~\mrm{cm}^{-3}$ is the atomic density of SiC. 

Positron trapping coefficients for negative vacancies at 300 K range typically between $0.5..5\times10^{15}~\mrm{s}^{-1}$ in semiconductors.~\cite{posspect} For instance, a value of $\mu_\mrm{V}^- = 1.1\times10^{15}~\mrm{s}^{-1}$ has been reported for $V_\mrm{Si}V_\mrm{C}$~\cite{henry_03} in SiC at room temperature.
Also, coefficient as high as $6\times 10^{16}~\mrm{s}^{-1}$ have been reported for $V_\mrm{Si}^{-}$ in  SiC.~\cite{kawasuso_98} We consider the latter value to be much too high, as the typical coefficients reported for semiconductors are an order-of-magnitude less.
It should be noted, however, that determining the values for the trapping coefficient $\mu$ is not an easy task and thus one should consider the possibility of the coefficient differing even by factor 2-3 from the physically proper value. Using an inaccurate trapping coefficient would thus affect the determined absolute defect concentrations (linearly). Still, the error would be similar in all samples and the comparison of the concentrations between different samples is possible.

Due to high uncertainties associated for the value of $\mu_V$, we use positron trapping coefficient at 300 K $\mu_\mrm{V}^- = 2\times10^{15}~\mrm{s}^{-1}$ for singly negative vacancies in Si sublattice, which is a typical value in wide band-gap semiconductors.\cite{posspect} 
For neutral vacancies we use $\mu_\mrm{V}^0 = 1\times10^{15}~\mrm{s}^{-1}$, since the trapping coefficients of the neutral and negative mono- and divacancies in Si have been found to differ by a factor of 1.5-3.5. \cite{mascher_89,makinen_92,hautojarvi_93} 
In n-type samples we use a value $3\mu_\mrm{V}^-$, since theoretical calculations predict that $V_\mrm{Si}$ changes it charge state from 1$^-$ to 2$^-$, and eventually to 3$^-$, when the Fermi level approaches the conduction band.~\cite{torpo_01} This change of charge state of $V_\mrm{Si}$ is likely to occur in the sample A2, where the conductivity of the sample changes from semi-insulating to $n$-type, indicating the movement of the Fermi level. 
No change in the conductivity of samples A1 and B was observed.
For small vacancy clusters (transition limited positron trapping), the positron trapping coefficient for vacancy clusters of $n$ vacancies can be approximated as $\mu_\mrm{CL} = n\mu_\mrm{V}$.~\cite{nieminen_79} As can be observed in Fig.~\ref{fig:LT_TEMP}, a part of the vacancy clusters in the annealed sample A1 are likely to be negative, which increases the overall trapping coefficient of the clusters. Hence the concentration of the vacancy clusters in the annealed sample A1 is probably a bit overestimated. The determined vacancy defect concentrations are summarized in Table~\ref{tab:samples_results}, being in the mid-$10^{16}$~cm$^{-3}$ for the Si vacancy related defects and $10^{15} - 10^{16}$~cm$^{-3}$ for the vacancy clusters.

The experimental vacancy defect concentrations and sizes (Table~\ref{tab:samples_results}) suggest the following interpretation:  Annealings of the samples of type A, grown under  hydrocarbon rich conditions,
increase the sizes of the vacancy clusters from approximately
5 to about 30 vacancies, with a simultaneous decrease of the cluster concentration. 
The Si vacancy complex concentration is also lowered
by the annealing of the n-type sample A1.
Thus we suggest that annealing causes $V_\mrm{Si}$ to (partly) agglomerate to the vacancy clusters by migration.

In the originally semi-insulating sample A2, the trapping rate $\kappa_1$ of positrons to the $V_\mrm{Si}$-related defects increases roughly by a factor of two. However, the annealing decreases the resistivity of the sample (towards $n$-type conductivity), which indicates that the Fermi level moves towards the conduction band. This is likely to induce an increase of the negativity of the $V_\mrm{Si}$, which would increase the trapping coefficient to these vacancies by a factor of 2--3, depending on the initial and final charge states. Hence the concentration of the $V_\mrm{Si}$-related defects decreases during the annealing.
However, it is worth noticing that the estimation of the concentration of the $V_\mrm{Si}$-related defects is less accurate than in the sample A1 due to the likely change in the charge state of the defects.
Interestingly, the total open volume increases in the annealing of the A-type sample. Based on these observations, we suggest that the annealing causes $V_\mrm{Si}$ to agglomerate to the clusters. 

Annealing increases the average size of the vacancy clusters also in the sample of type B, grown under hydrocarbon poor conditions. The concentration of the $V_\mrm{Si}$-related defects is not significantly changed in the annealing.



\subsection{Electrical compensation}

The positron results lead to several important conclusions regarding the origin of the semi-insulating properties of the material.
The temperature dependence of the average positron lifetime  (Fig.~\ref{fig:LT_TEMP}) shows that the samples of both types A and B do not contain significant concentrations of negative ions, such as impurities or negative interstitials or antisite defects.
These types of defects thus contribute little to the electrical compensation,
at maximum at the level of their detection limit; their concentration is at most in the mid-$10^{15}~\mrm{cm}^{-3}$ range.
This observation correlates also with low concentrations of B and Al acceptors ($<5\times10^{15}~\mathrm{cm}^{-3}$  and $<5\times10^{14}~\mathrm{cm}^{-3}$, respectively) measured with secondary ion-mass spectrometry.

\begin{figure}[tb]
\centering{\includegraphics[angle=0,width=0.95\linewidth]{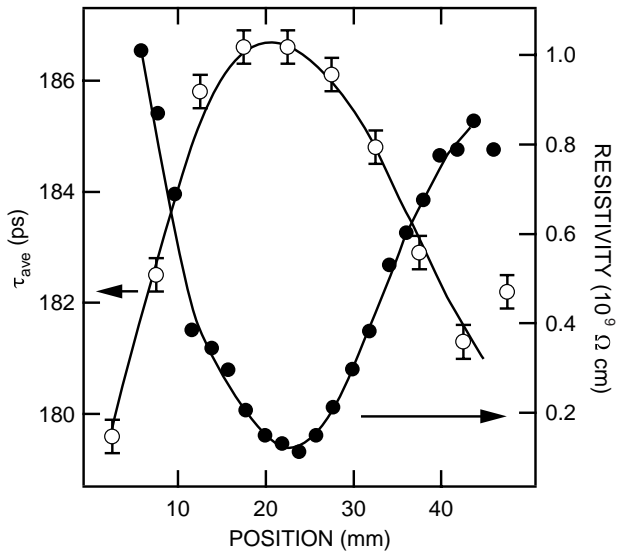}}
\caption{Average positron lifetime and resistivity as a function of position on the diameter of a 2'' type A wafer. The lines are for guiding the eye.}
\label{fig:LT_RESIST}
\end{figure}

The samples grown in hydrocarbon rich conditions (type A) are either slightly n-type already after the growth, or lose their high resistivity after annealing. The detected negative $V_\mrm{Si}$-related complex is an obvious candidate for the compensating intrinsic defect in as-grown material. Hence the n-type character of the annealed samples can be explained by the loss of compensation due to migration of negative $V_\mrm{Si}$ to primarily neutral vacancy clusters. It is worth noticing that a part of the vacancy clusters turn negative in the annealing of the sample A1. This does not affect the interpretation that the compensation is weakened due to the migration of $V_\mrm{Si}$ to the clusters, however, as the negativity implies that the cluster concentration presented in Table~\ref{tab:samples_results} overestimates the ``proper'' value.

The concentration of the vacancy clusters should thus increase with the decreasing resistivity of HTCVD SiC.  To verify this relation in the as-grown samples, we measured positron lifetime and resistivity  as a function of position in a strip like sample, cut diagonally across a wafer grown in hydrocarbon rich growth conditions (type A).  The lifetime component $\tau_\mrm{2} = 247\pm10$ ps remained constant over the sample, indicating constant vacancy cluster size ($\approx 4$ atoms).  As can be seen in  Fig.~\ref{fig:LT_RESIST}, the average positron lifetime and the resistivity of the sample anticorrelate. The change of $\tau_\mrm{ave}$ is caused by the variation in the intensity $I_2$ and thus the positron trapping rate at the vacancy clusters.
The concentration of the clusters thus decreases with increasing resistivity of the material, most likely since less compensating $V_\mrm{Si}$-related complexes have migrated to the neutral vacancy clusters during the growth.
Unfortunately, the direct analysis of the concentrations of  $V_\mrm{Si}$-related defects using eqs.~\ref{eq:kappa1} and \ref{eq:kappa2} is not possible in the case of the data of Fig.~\ref{fig:LT_RESIST}, since (i) the changes of the average lifetime are small, only about 6 ps, (ii) the single-trap model is almost valid (\emph{i.e.} the concentration of the $V_\mrm{Si}$ related defects is barely above the sensity limit $10^{16}~\mrm{cm}^{-3}$) and (iii) the differing position of the Fermi-level may have an influence on the charge state of $V_\mrm{Si}$-related defects.

The annealing of samples grown under hydrocarbon poor conditions (type B) leads to growth of the vacancy clusters, but the concentration of the smaller defects is not substantially changed. This suggests that in B-type material the Si vacancies have transformed (or belong to) to more stable complexes, such as the $V_\mrm{C}\mrm{C}_\mrm{Si}$~\cite{lingner_PRB_01,bockstedte_PRB_04}, or divacancies~\cite{son_PRL_06}, which remain as compensating centers after the annealing.  It is also possible that the high resistivity of type B samples could also be dominated by the C vacancy,~\cite{zvanut,konovalov,SonECSCRM2002,son_MSF_04,magnusson_MSF_06} which is more stable than $V_\mrm{Si}$ according to calculations.~\cite{rauls_03,bockstedte_PRB_04}

\section{Conclusions} \label{sec:conclusions}

Positron lifetime measurements in semi-insulating HTCVD SiC samples reveal vacancy clusters of the size of 4--5 missing atoms. Their charge state in the semi-insulating material is neutral. The clusters grow in size to about 30 missing atoms in post-growth annealings at 1600 $^\circ$C. Positron experiments identify also Si vacancy related smaller complexes, which are negatively charged. They partly disappear during annealing in materials grown under hydrocarbon rich conditions. After annealing the materials become more $n$-type, as part of the compensation is lost by the migration of Si vacancy related defects to (partially) increase the size of the open volume of clusters. Hence, our results suggest that the Si vacancy complexes act as compensating centers partly responsible for the insulating properties of high purity HTCVD SiC. This conclusion is supported by the measurement of resistivity variations across a SiC wafer, which shows that concentration of the vacancy clusters anticorrelates with the resistivity.

\begin{acknowledgements}
One of the authors (R.A.) gratefully acknowledges the support from Finnish Academy of Science and Letters, Vilho, Yrjö and Kalle Väisälä Foundation.
\end{acknowledgements}

\bibliography{SiCBib}

\end{document}